\documentclass[preprint,amsmath,amssymb,aps,longbibliography]{revtex4-2}
\usepackage{lineno,hyperref}
\usepackage{bm}
\usepackage{amsmath}
\usepackage{gensymb}
\usepackage{epstopdf}
\usepackage{booktabs}
\usepackage{float}
\usepackage{multirow}

\usepackage{graphicx}
\usepackage{xcolor}
\usepackage{soul}
\begin{document}

\title{Randomness in atomic disorder and consequent squandering of spin-polarization in a ferromagnetically fragile quaternary Heusler alloy FeRuCrSi}

\author{Shuvankar Gupta$^{1}$}
\email{guptashuvankar5@gmail.com}
\author{Sudip Chakraborty$^{1}$}
\author{Vidha Bhasin$^{2}$}
\author{Celine Barreteau$^3$}
\author{Jean-Claude Crivello$^{3,4}$}
\author{Jean-Marc Greneche$^5$}
\author{S.N. Jha$^6$}
\author{D. Bhattacharyya$^2$}
\author{Eric Alleno$^3$}
\author{Chandan Mazumdar$^{1}$}
\email{chandan.mazumdar@saha.ac.in}

\affiliation{$^1$Condensed Matter Physics Division, Saha Institute of Nuclear Physics, 1/AF, Bidhannagar, Kolkata 700064, India}
\affiliation{$^2$Atomic \& Molecular Physics Division, Bhabha Atomic Research Centre, Mumbai 400 094, India}
\affiliation{$^3$Univ Paris Est Creteil, CNRS, ICMPE, UMR 7182, 2 rue H. Dunant, 94320 Thiais, France}
\affiliation{$^4$CNRS-Saint-Gobain-NIMS, IRL 3629, Laboratory for Innovative Key Materials and Structures (LINK), 1-1 Namiki, 305-0044 Tsukuba, Japan}
\affiliation{$^5$Institut des Mol\'{e}cules et Mat\'{e}riaux du Mans, IMMM, UMR CNRS 6283, Le Mans Universit\'{e}, Avenue Olivier Messiaen, Le Mans Cedex 9, 72085, France}
\affiliation{$^6$Beamline Development and Application Section Physics Group, Bhabha Atomic Research Centre, Mumbai 400085, India}
\date{\today}
\begin{abstract}
The Fe$_x$Ru$_{2-x}$CrSi ($0<x<2$) system should theoretically reckoned to be one of the few known examples of a robust half-metallic ferromagnet with 100\% spin polarisation, with Cr atoms deemed to be the main contributor to magnetism. In this system, the FeRuCrSi equiatomic member is considered a technologically important material, since band structure calculations suggest that it is a spin-gap-free semiconductor. Through our extensive structural analysis of FeRuCrSi using X-ray diffraction, extended X-ray absorption fine structure and $^{57}$Fe M\"{o}ssbauer spectrometry, we have found a random disorder between the Fe and Ru sites. Contrary to theoretical predictions, the magnetic moment in this system is in fact contributed by Fe atoms, which calls into question the fundamental basis of the half-metallic character proposed for the Fe$_x$Ru$_{2-x}$CrSi series. Our M\"{o}ssbauer results also reveal a rare scenario in which key physical properties are intricately correlated with material chemistry in the form of random atomic disorder on a localised scale.
\end{abstract}
\maketitle

\section{\label{sec:Introduction}Introduction}

Spintronics has emerged as a promising field in nanoelectronics, offering the potential for more efficient and powerful devices by mitigating power consumption while enhancing memory and processing capabilities \cite{awschalom2007challenges,vzutic2004spintronics,pulizzi2012spintronics}. 
A critical aspect of spintronics is the exploration of materials with high spin-polarization, deemed ideal for advancing spintronic technologies \cite{felser2007spintronics}. 
Particularly, half-metallic ferromagnets (HMFs) \cite{de1983new} and spin gapless semiconductors (SGSs) \cite{wang2008proposal} are recognized for their high spin polarization and versatile magnetic properties, making them valuable for enhancing the efficiency of spintronic devices across various applications such as spin valves \cite{ikhtiar2014magneto}, data storage, magnetic sensors \cite{felser2007spintronics}, spin injectors \cite{saito2013spin}, and magnetic tunnel junctions \cite{kubota2009half}.

Heusler alloys have recently garnered attention due to their diverse properties, including half-metallicity \cite{alijani2011quaternary,PhysRevB.84.224416,gupta2022coexisting,PhysRevB.108.045137,D3TC02813E}, spin gapless semiconductivity \cite{bainsla2015spin,bainsla2015origin,D3TC03481J}, spin-semimetals \cite{venkateswara2019coexistence,venkateswara2023ferhcrsi}, spin glass \cite{PhysRevB.107.184408}, re-entrant spin glass \cite{PhysRevB.108.054405,PhysRevB.108.054430,chakraborty2024rare}, bipolar magnetic semiconducting characteristics~\cite{nag2021bipolar} and Curie temperature~\cite{PhysRevB.108.245151}, \textit{etc}. 
The exploration of new Heusler alloys is crucial for advancing this area of study, as despite theoretical predictions~\cite{xu2013new,gao2019high,skaftouros2013search,gao2015first} of numerous Heusler alloys possessing SGS features, only a limited number of experimental realizations have been achieved so far~\cite{ouardi2013realization,bainsla2015spin,bainsla2015origin,D3TC03481J}. 
Among them, Mn$_2$CoAl \cite{ouardi2013realization}, CoFeMnSi \cite{bainsla2015spin}, CoFeCrGa \cite{bainsla2015origin}, and CoFeMnSn \cite{D3TC03481J} have been experimentally established as SGS materials, with the additional discovery of the fully compensated ferrimagnetic SGS CrVTiAl \cite{PhysRevB.97.054407}. 
It is pertinent to note here that the most of the Heusler alloys often form with inherent and irrepressible structural disorder which is generally argued to be responsible for the nonconformity with  the theoretically predicted HMF and SGS properties. Thus, the search for a compound with HMF and SGS properties, resistant to disorder, will be an effective solution for this type of situation.

Fe$_x$Ru$_{2-x}$CrSi is one such system where first principles band structure calculations predict robust HMF properties against structural disorder throughout the series~\cite{mizutani2006half,hiroi2009magnetic,PhysRevB.79.224423}. 
The equiatomic member of the series, FeRuCrSi, has been additionally predicted to exhibit SGS characteristics~\cite{guo2018magnetic,zhang2023atomic,wang2017structural}. 
The quaternary compound FeRuCrSi can be thought of as the intermediate between two ternary Heusler alloys Fe$_2$CrSi~\cite{luo2007electronic} and Ru$_2$CrSi~\cite{hiroi2013antiferromagnetic}. 
Since Fe and Ru are isoelectronic atoms, the total valence electron count (VEC) of all members of Fe$_x$Ru$_{2-x}$CrSi remain invariant (VEC = 26) irrespective to the value of $x$. Following the Slater-Pauling (S-P) rule~\cite{galanakis2002slater,ozdougan2013slater} for HMF Heusler alloy, the total magnetic moment for any member of the series is expected to be $M$ = (VEC-24) = 2\,$\mu_B$/f.u. 
However, the Ru-rich end-composition, Ru$_2$CrSi is experimentally found to undergo antiferromagnetic transition below $T_N\sim$14\,K~\cite{hiroi2013antiferromagnetic}. 
Subsequent theoretical calculations indeed show lower ground state energy for the antiferromagnetic spin arrangement in this compound~\cite{bahlouli2013structure}. 
As Fe concentration is gradually increased, the system initially transforms to a spin-glass state, before it finally settles to a ferromagnetic ground state for $x>0.3$~\cite{matsuda2005magnetic,hiroi2021high,hiroi2013muon,PhysRevB.79.224423,ito2010low}. Fe$_2$CrSi, the other-end member of the series, exhibits a Curie temperature at 520\,K with saturation magnetic moment of 2\,$\mu_B$/f.u., in line with the S-P rule~\cite{luo2007electronic}. 
Theoretical calculations suggest Cr-atoms are the major contributor of magnetism in this series of compounds, although this postulation is yet to be experimentally verified  for compositions with  $x\neq 0$. 
Interestingly, some of the experimental reports rather suggest that a linear variation of the magnetic moment occur with Fe substitution in Fe$_x$Ru$_{2-x}$CrSi, contrary to the invariance of the magnetic moment expected from the S-P rule for HMF systems~\cite{matsuda2005magnetic}. Additionally, even for FeRuCrSi, substantial variations in its basic characteristics like Curie temperature ($T_{\rm C}$) and saturation magnetization ($M_{sat}$) at low temperature have been reported in the literature~\cite{matsuda2005magnetic,zhang2023atomic}. 
Thus, the members of Fe$_x$Ru$_{2-x}$CrSi series not only sometimes fail to conform to the theoretical prediction, the experimental results on $T_{\rm C}$ and $M_{Sat}$ reported by different groups on the same compositions are also not reproducible. 
As mentioned earlier, since the Heusler alloys are known to form with inherent structural disorder, \textit{a priori} one would expect that such variations in properties reported by various groups are related to different natures and extents of disorder present in the respective materials, despite having the same nominal composition. 
In this work, we have focussed on the equiatomic FeRuCrSi, and investigated the nature of disorders to explain the above-mentioned discrepancies among different experimental results as well as between experiment and theory. 
Very surprisingly, our investigation reveals that completely contrary to the theoretically established magnetic moment on Cr in FeRuCrSi, the magnetism is actually contributed by Fe atoms. The reported variations in $T_{\rm C}$ and $M_{sat}$ can also be explained by the local arrangement of Fe/Ru in each unit cells. Our work thus firmly establishes not only the role of anti-site disorders, but the effect of randomness of Fe/Ru distribution in this compound as well.

\section{Methods}

\subsection{Experimental}
The polycrystalline FeRuCrSi compound was synthesized via arc melting, employing high-purity constituent elements (purity $>$ 99.99\%), under argon atmosphere. The synthesis process involved 5-6 melting cycles, with flipping the sample between each cycle to ensure composition homogeneity. A part of the sample was annealed for 5 days at 1073\,K, followed by a quench in an ice-water mixture. Room temperature powder X-ray diffraction (XRD) analysis was performed using Cu-K$\alpha$ radiation on a TTRAX-III diffractometer (Rigaku, Japan). The single-phase composition and crystal structure of the sample were determined through Rietveld refinement using the FULLPROF software package~\cite{rodriguez1993recent}.

Extended X-ray absorption fine structure (EXAFS) measurements were conducted at the Energy Scanning EXAFS beamline (BL-9) at the Raja Ramanna Centre for Advanced Technology (RRCAT) in Indore, India, employing standard EXAFS measurement protocols~\cite{PhysRevB.108.045137,PhysRevB.108.054405}. Magnetic measurements were carried out in the temperature range of 3--380\,K under magnetic fields up to 70 kOe using a commercial SQUID-VSM instrument (Quantum Design Inc., USA).
${^{57}}$Fe transmission M\"{o}ssbauer spectrometry was conducted at many different temperatures in the range of 405-77 K to investigate the local atomic environments of Fe atoms in the sample. An additional measurement under an externally applied magnetic field of 0.06 Tesla has also been carried out to investigate the influence of magnetic field in the M\"{o}ssbauer spectra. The samples consisted of a thin powder layer containing about 5 mg Fe/cm$^2$. Spectra were collected using an electromagnetic transducer with a triangular velocity profile and a ${^{57}}$Co source diffused into a Rh matrix within a bath cryostat for measurements below 300 K. The hyperfine structures were modelled using a least-square fitting procedure, involving quadrupolar doublets composed of Lorentzian lines, utilizing the in-house program `MOSFIT'. Isomer shift values were reported with reference to ${\alpha}$--Fe at 300\,K, and the velocity was controlled using a standard of ${\alpha}$--Fe foil.

\subsection{Computational}

The enthalpy of formation, electronic structure, and spin polarization at ground state were determined using Density Functional Theory (DFT) calculations with the projector augmented wave (PAW) method \cite{blochl1994PAW}, as implemented in the Vienna ab initio simulation package (VASP) \cite{kresse1993vasp1, kresse1994vasp2}. The exchange-correlation was modelled by the generalized gradient approximation modified by Perdew, Burke, and Ernzerhof (GGA-PBE)~\cite{perdew1996gga}. Detailed information on calculation parameters can be found in previous articles~\cite{gupta2022coexisting,PhysRevB.108.045137,D3TC02813E,PhysRevB.108.054405}.

\section{Results and Discussion}

\subsection{\label{sec:Electronic_Structure_Ordered} Structure optimization and electronic structure calculations}

\begin{figure*}[]
\centerline{\includegraphics[width=.98\textwidth]{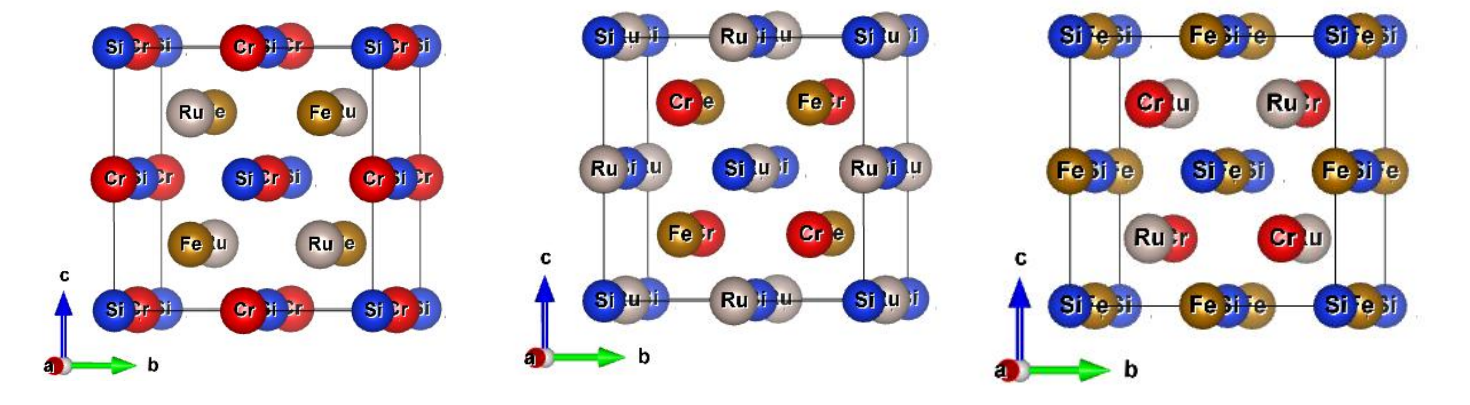}}
{\caption{Atomic arrangements of FeRuCrSi for (a) Type-1 (b) Type-2 (c) and Type-3 structure.}\label{Structure_Ordered}}
\end{figure*}

\begin{figure*}[]
\centerline{\includegraphics[width=.98\textwidth]{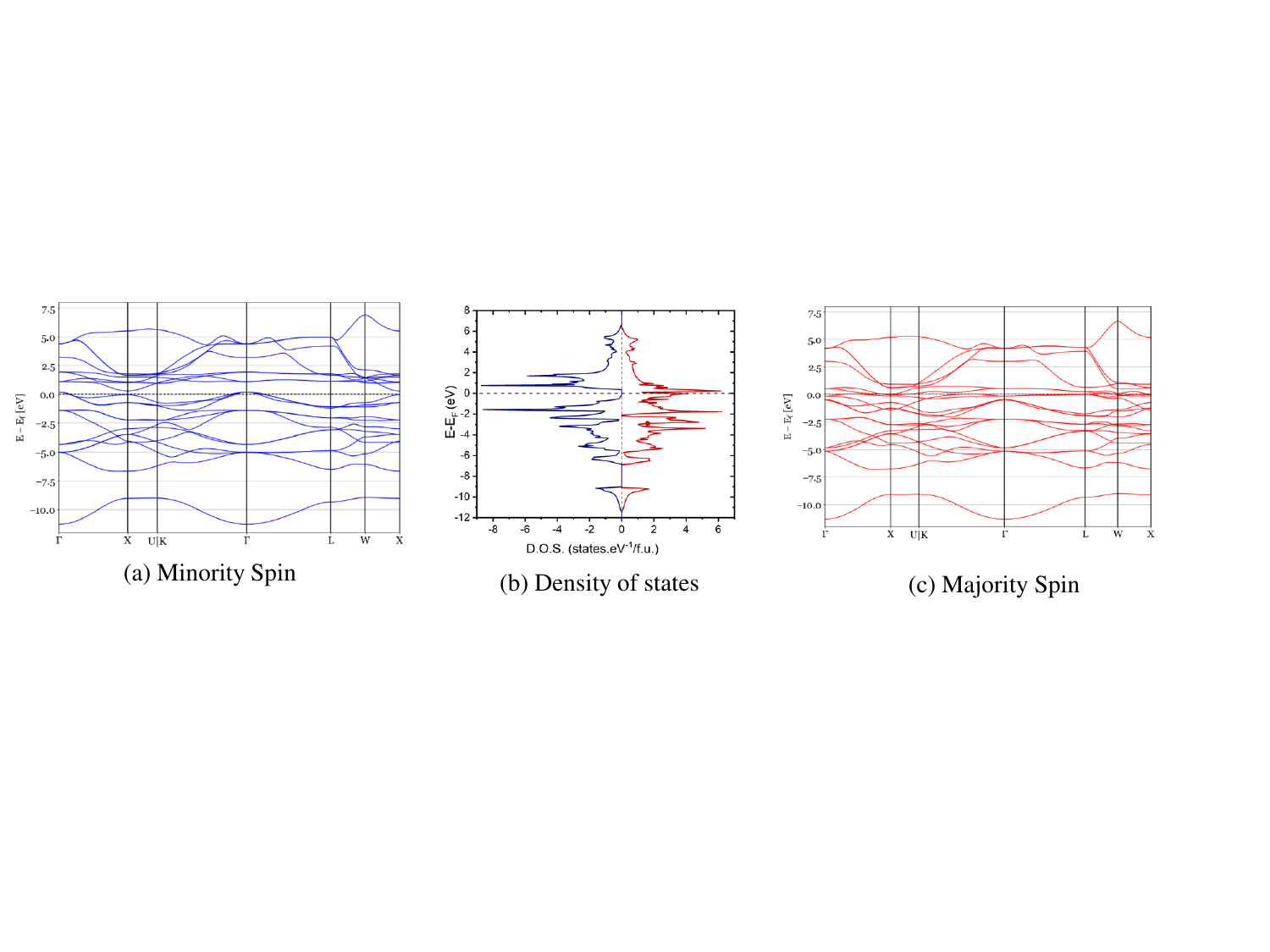}}
{\caption{Spin-polarized band structure and density of states of FeRuCrSi in ordered Type-1 structure: (a) minority spin band (b) density of states, (c) majority spin band.}\label{DOS}}
\end{figure*}

\begin{table*}[ht]
\addtolength{\tabcolsep}{3.0pt}
\caption{Calculated enthalpy of formation (${\Delta_f{H}}$) and magnetic moment for 3 different ordered atomic arrangements of FeRuCrSi and 2 different disordered structures. For these laters, the magnetic moment on $4c$ and $4$ corresponds to the average value of atoms in the respective ordered structure, e.g. Ru in $4c$ for Type-1.}
\label{Enthalpy}
\begin{tabular}{c|c|c|c|c|c|c|c|c|c|c}
\hline
 & \multicolumn{4}{c|}{sites distributions} & $\Delta_f{H}$ & \multicolumn{5}{c}{magnetic moment ($\mu_B$)} \\
 & 4$a$ & 4$b$ & 4$c$ & 4$d$ & (kJ/mol by atom) & 4$a$ & 4$b$ & 4$c$ & 4$d$ & Tot \\
\hline 
\hline
Type-1 ordered & Si & Cr & Ru & Fe & -33.63 & 0.00 & 1.98 & -0.18 & 0.15 & 1.94 \\
\hline
Type-1 disordered & Si & Cr & \multicolumn{2}{c|}{(Ru,Fe)} & -34.13 & 0.00 & 1.90 & -0.16 & 0.21 & 1.94\\
\hline
Type-2 ordered & Si & Ru & Cr & Fe & -4.04 & 0.00 & 0.01 & 0.00 & 0.01 & 0.02 \\
\hline
Type-3 disordered  & Si & Fe & Cr & Ru & -29.25 & 0.04 & 2.56 & -1.18 & 0.30 & 1.73 \\
\hline
Type-3 ordered & Si & Fe & \multicolumn{2}{c|}{(Cr,Ru)} & -29.61 & 0.05 & 2.46 & -0.87 & 0.19 & 1.85\\
\hline
\hline
\end{tabular}
\end{table*}

A quaternary Heusler alloy, denoted as ${XX'YZ}$, possesses four distinct crystallographic sites in space group $F\overline{4}3m$: 4\textit{a} (0,0,0), 4\textit{b} (0.5,0.5,0.5), 4\textit{c} (0.25,0.25,0.25), and 4\textit{d} (0.75,0.75,0.75). Generally, the \textit{sp}-element \textit{Z} occupies the 4\textit{a} position, while the transition metals reside in the remaining three sites. 
Although there are six possible configurations, permutations of atoms in the 4\textit{c} and 4\textit{d} positions result in energetically equivalent configurations. 
Consequently, only the three distinct configurations (Type-1, Type-2, and Type-3) were considered, as shown in Fig.\ref{Structure_Ordered} and Table.\ref{Enthalpy}. The relative stabilities of these configurations were assessed through DFT calculations by comparing their enthalpies of formation.

Our DFT calculations on the ordered structure revealed that the Type-1 configuration, with Si at 4\textit{a}, Cr at 4\textit{b}, Ru at 4\textit{c}, and Fe at 4\textit{d}, is the most stable ordered configuration expected at 0\,K. 
The spin-polarized band structure and density of states (DOS) for the Type-1 configuration are depicted in Fig.~\ref{DOS}. The DOS illustrates a band gap at the Fermi level (\textit{$E_{\rm F}$}) for the spin-down band, while the spin-up band is indicative of a metallic character. 
This unique band structure characterizes the material as a Half-Metallic Ferromagnetic (HMF) compound, as affirmed by a very high polarization of $P=\frac{\rm{DOS}^\uparrow (E_{\rm F})- \rm{DOS}^\downarrow (E_{\rm F})}{\rm{DOS}^\uparrow (E_{\rm F})+ \rm{DOS}^\downarrow (E_{\rm F})}$ = 100$\%$. Our result thus matches well with the theoretically calculated DOS on the same compound predicting the HMF character~\cite{wang2017structural}, although some other studies additionally predicted SGS characteristics as well~\cite{guo2018magnetic,zhang2023atomic}.

Consistent to that expected for HMF systems with VEC 26~\cite{galanakis2002slater,galanakis2006electronic,ozdougan2013slater}, our calculation also yields the total magnetic moment to be 1.94\,${\mu_B}$. Moreover, atom-specific magnetic moments were computed and reported in Table.~\ref{Enthalpy}, and are in good agreement with earlier published reports~\cite{wang2017structural,guo2018magnetic,zhang2023atomic}. The theoretical analysis suggests that Cr significantly contributes to the magnetic moment, and the moment on the Fe site rather exhibits an induced character. Similar behaviour has earlier been observed in FeMnVAl ~\cite{gupta2022coexisting} and FeMnVGa ~\cite{PhysRevB.108.045137} where magnetic atom Mn and nonmagnetic Fe are distributed randomly in the 4\textit{c} and 4\textit{d} sites, although in FeRuCrSi, Cr would occupy 4\textit{b} sites whereas Ru and Fe would occupy 4\textit{c} and 4\textit{d} positions, respectively.

\subsection{\label{sec:XRD}X-ray diffraction}
\begin{figure}[h]
\centerline{\includegraphics[width=.48\textwidth]{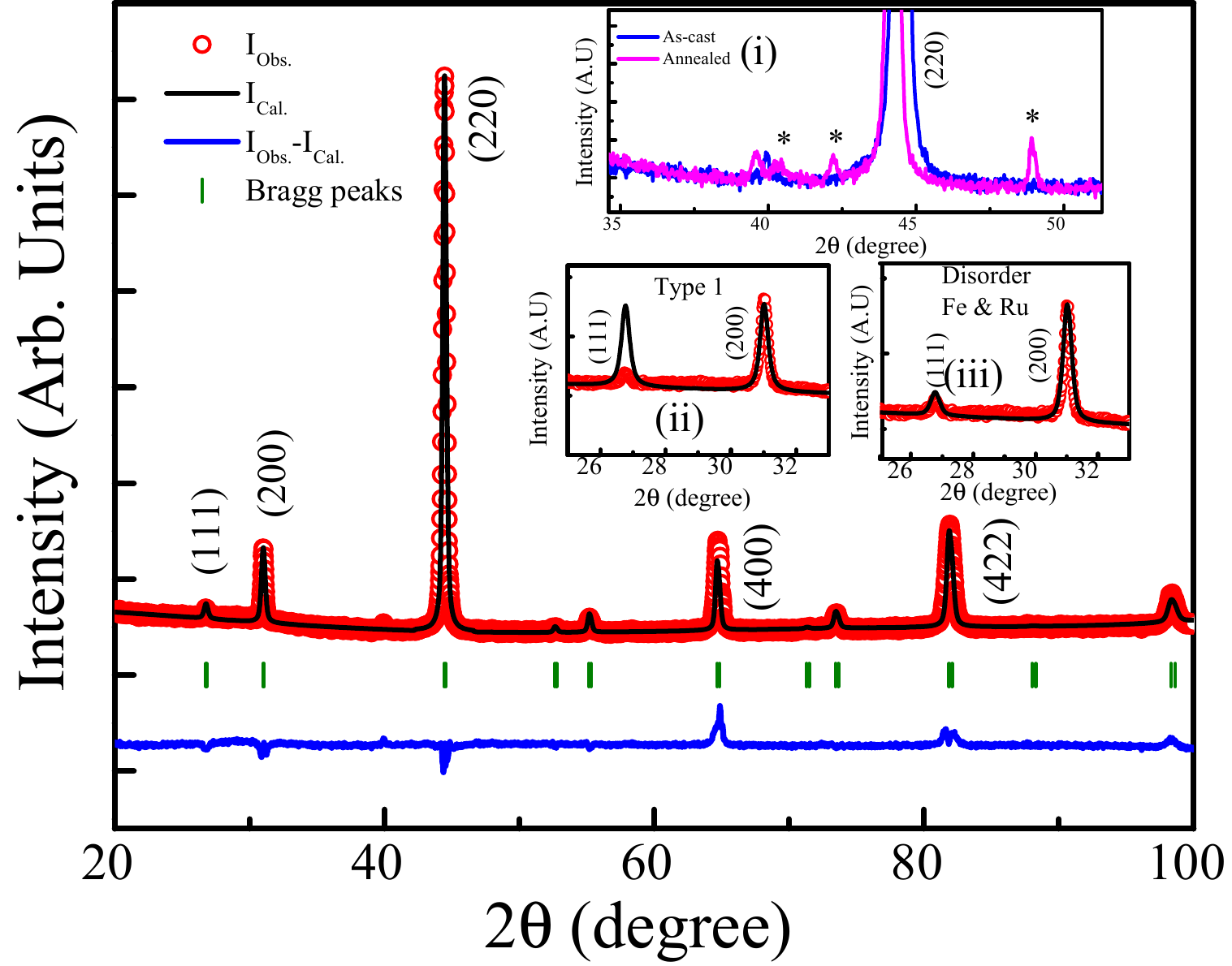}}
{\caption{Rietveld refinement of the powder XRD pattern of FeRuCrSi (disordered structure) measured at room temperature. Inset (i) shows the magnified view of the XRD data for as-cast and annealed compound, where the impurity peaks in the annealed sample are marked by asterisk sign. Insets (ii) and (iii) show the magnified view of the Rietveld refinement results considering the ordered structure and disordered structure, respectively.}\label{XRD}}
\end{figure}

We have carried out XRD measurements at room temperature for both as-cast as well as annealed sample. We found a few additional small peaks in the XRD patterns of the annealed sample which are absent in the XRD pattern of the as cast samples and also not conforming to the cubic character of Heusler alloy (inset (i) of Fig.~\ref{XRD}). The result thus suggest that annealing tend to introduce secondary phase in the system. Consequently, all further measurements are carried out on the as-cast compound only. 
The Fig.~\ref{XRD} represent the Rietveld refinement of the XRD pattern of as-cast FeRuCrSi. Theoretical calculations suggests that ordered Type-1 structure possess minimum energy compared to other ordered structures. However, the earlier published experimental results on the crystal structure of FeRuCrSi are quite contradictory. Although some of the reports suggested the ordered LiMgPdSn structure-type~\cite{matsuda2005magnetic}, a few others also reported disordered atomic arrangements without determining the exact nature of disorder~\cite{zhang2023atomic}.

Our attempt to fit the XRD pattern using the ordered Type-1 structure model however failed to properly address the superlattice (111) Bragg peak intensity (inset (ii) of Fig.~\ref{XRD}). For Heusler systems, the (111) and (200) Bragg peak intensities are considered to be indicative of structural disorder present in the system. For example, in \textit{A2}-type (random mixing of elements in 4\textit{a}, 4\textit{b}, 4\textit{c} and 4\textit{d} sites) disorder, both the peaks are found to be absent whereas in \textit{B2}-type (cross-substitution between 4\textit{a} \& 4\textit{b} and between 4\textit{c} \& 4\textit{d}) only the (111) peak vanishes. In our case, both the 111 and 200 lines are present, thus ruling out \textit{A2} and \textit{B2}-type disorder (inset (iii) of Fig.~\ref{XRD}). Nonetheless, the intensity of 111 line is substantially reduced when compared to the theoretically estimated one for an ordered Type-1 structure. The reduction in intensity of the (111) peak, however, could be explained if the disorder is limited to the 4\textit{c} and 4\textit{d} sites only, where Fe and Ru atoms are distributed in 50:50 ratio (Type-1 disorder structure), averaging over the sample volume.
However, it may be noted that the above structural model arising out of the XRD analysis, although is the simplest, yet certainly not unique. For example, a more complex structural model involving a combinations of Type-1 and Type-3 disorder with Si at 4\textit{a}, Cr$_{0.5}$Fe$_{0.5}$ at 4\textit{b}, Ru$_{0.5}$Fe$_{0.25}$Cr$_{0.25}$ at 4\textit{c}  and Fe$_{0.25}$Cr$_{0.25}$Ru$_{0.5}$ at 4\textit{d} can also explain the XRD spectra with almost equal satisfaction. Nevertheless, as many other quaternary Heusler alloys, e.g.  CoRuMnSi~\cite{venkateswara2020half}, FeMnVAl~\cite{gupta2022coexisting}, FeMnVGa~\cite{PhysRevB.108.045137}, NiRuMnSn~\cite{PhysRevB.108.054405} and FeRuVSi~\cite{D4TC02267J}, etc. exhibiting similar XRD patterns have been established to form in the simpler disordered crystal structure by various theoretical and experimental tools, including neutron diffraction, we believe the same would also hold true for FeRuCrSi as well. The magnetic field dependent $^{57}$Fe M\"{o}sbauer spectroscopic data analysis of FeRuCrSi, presented later in Sec.~\ref{sec:Mossbauer}, indicates extremely local variation of hyperfine field which is more in consonance with the atomic disorder within the Type-I structure, rather than having a mixture of Type-I and Type-3 structure of much larger individual cluster sizes. The lattice parameter is found to be 5.757 \,{\AA}, and matches with most of the other reported values~\cite{matsuda2005magnetic,wang2017structural}.

\subsection{\label{sec:EXAFS}EXAFS}

\begin{figure}
\begin{minipage}{0.49\textwidth}
\centering
{\includegraphics[width=.98\textwidth]{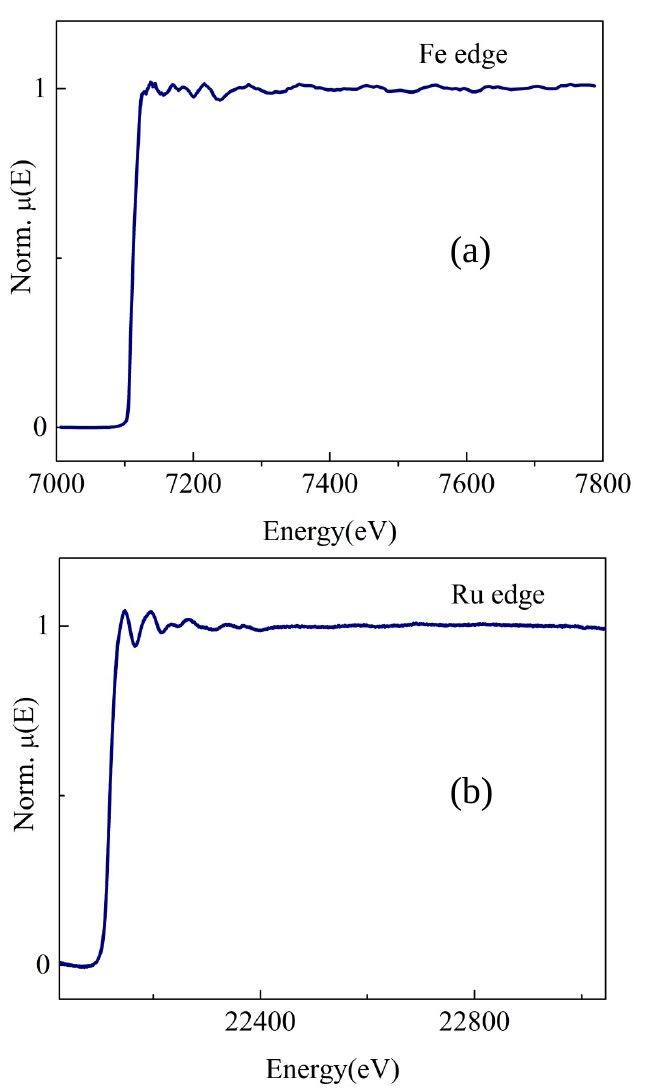}}
{\caption{Normalized EXAFS spectra of FeRuCrSi taken at (a) Fe edge and (b) Ru edge}\label{EXAFS_Data}}
\end{minipage}
\end{figure}

\begin{figure}
\begin{minipage}{0.49\textwidth}
\centering
{\includegraphics[width=.98\textwidth]{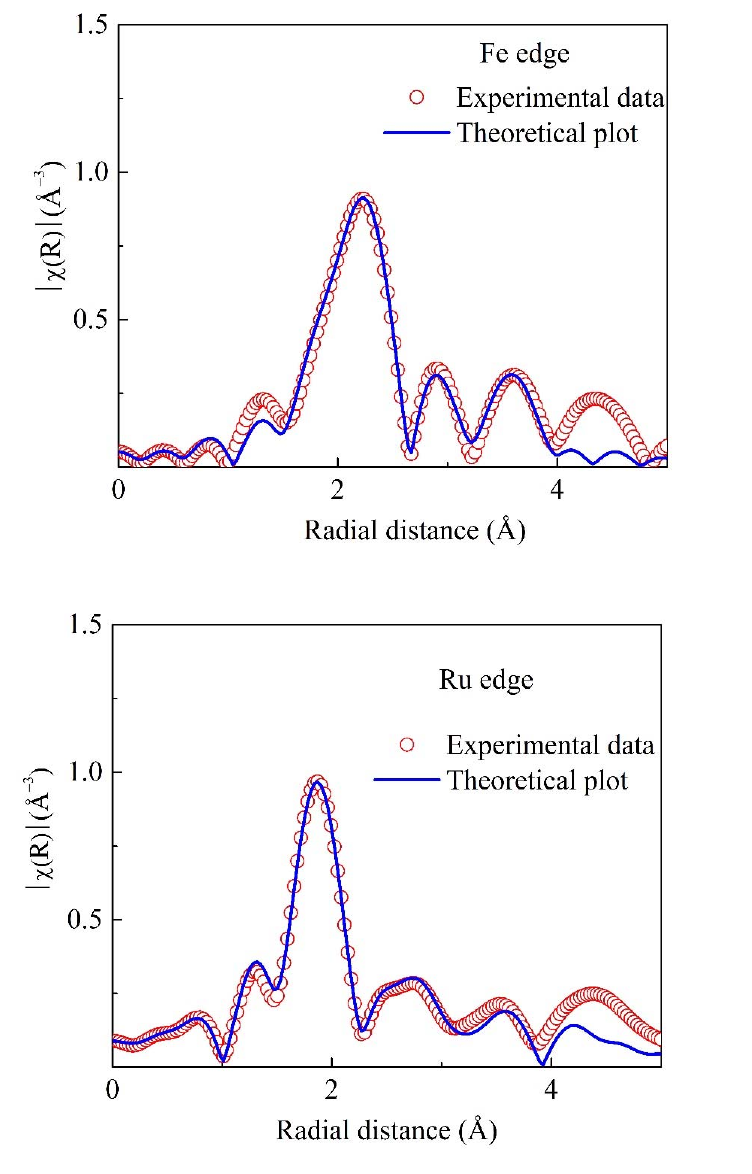}}
{\caption{Fourier transformed EXAFS spectra of FeRuCrSi taken at (a) Fe edge and (b) Ru edge.}\label{EXAFS_Fitted}}
\end{minipage}
\end{figure}

\begin{table*}[ht]
\addtolength{\tabcolsep}{3.0pt}
\caption{Bond length (R), coordination number (N), and Debye-Waller or disorder factor (${\sigma}^2$) obtained by simultaneous fitting of EXAFS data of FeRuCrSi at Fe and Ru edges.}
\label{Tab:EXAFS}
\begin{tabular}{cccccccc}
\hline
\multicolumn{4}{c}{Fe edge} & \multicolumn{4}{c}{Ru Edge}  \\   \hline\hline
Path  & R ({\AA})   & N   & ${\sigma}^2$     & Path  & R ({\AA})   & N   & ${\sigma}^2$     \\\hline
Fe-Si & 2.48${\pm}$0.03 & 4   & 0.0300${\pm}$0.0055 & Ru-Si & 2.43${\pm}$0.08 & 4   & 0.0062${\pm}$0.0005 \\\hline
Fe-Cr & 2.49${\pm}$0.01 & 4   & 0.0089${\pm}$0.0005 & Ru-Cr & 2.50${\pm}$0.01 & 4   & 0.0072${\pm}$0.0007 \\\hline
Fe-Ru & 2.76${\pm}$0.01 & 3.5 & 0.0089${\pm}$0.0018 & Ru-Fe & 2.75${\pm}$0.01 & 3.3 & 0.0085${\pm}$0.0013 \\\hline
Fe-Fe & 2.75${\pm}$0.01 & 2.5 & 0.0018${\pm}$0.0010 & Ru-Ru & 2.79${\pm}$0.01 & 2.7 & 0.0092${\pm}$0.0016 \\\hline
Fe-Ru & 4.05${\pm}$0.01 & 6   & 0.0300${\pm}$0.0130 & Ru-Fe & 4.05${\pm}$0.03 & 6   & 0.0190${\pm}$0.0021 \\\hline
Fe-Fe & 4.05${\pm}$0.01 & 6   & 0.0160${\pm}$0.0020 & Ru-Ru & 4.13${\pm}$0.04   & 6   & 0.0175${\pm}$0.0028   \\\hline\hline
\end{tabular}
\end{table*}

To confirm the anti-site disorder between Fe and Ru atoms in FeRuCrSi, we have also carried out Extended X-ray Absorption Fine Structure (EXAFS) at Fe and Ru edges. Unlike XRD, EXAFS is an element-specific measurement that focuses on the local atomic environment surrounding the examined atoms. In the case of Heusler alloys, EXAFS has proven to be a valuable tool for elucidating precise atomic arrangements~\cite{balke2007structural,PhysRevB.108.045137,PhysRevB.108.054405}.

The normalized EXAFS ($\mu$(E) versus E) spectra measured at the Fe and Ru edges are illustrated in Fig.~\ref{EXAFS_Data}. The standard procedure was followed in processing the EXAFS data analysis~\cite{koningsberger1987x}. In brief, the absorption spectra ($\mu$(E) vs. \textit{E}) were transformed into the absorption function $\chi$(E), as given by eqn.~\ref{eq1}, to obtain quantitative information on the local structure.

\begin{equation}
\chi(E) = \frac{\mu(E)-\mu_{0}(E)}{\Delta\mu_{0}(E_{0})}
\label{eq1}
\end{equation}

The energy-dependent absorption coefficient $\chi$(\textit{E}) was then converted to the wave number-dependent absorption coefficient using eqn.~\ref{eq2}:

\begin{equation}
K=\sqrt{\frac{2m(E-E_{0})}{\hbar^{2}}}
\label{eq2}
\end{equation}

To enhance oscillations at high \textit{k}, $\chi(k)$ was weighted by \textit{k$^2$}. The resulting ${\chi(k)k^2}$ functions were Fourier-transformed in R space to produce $\chi$(R) versus R plots, representing actual distances from the center of the absorbing atom. The data reduction, including background reduction and Fourier transformation, were carried out utilizing the ATHENA subroutine within the Demeter software package~\cite{ravel2005athena}. The FeRuCrSi sample's Fourier-transformed EXAFS spectra at the Fe and Ru edges are presented in Fig.~\ref{EXAFS_Fitted} as $\chi$(R) versus R plots.

The goodness of fit was assessed using the R$_{factor}$ defined by Equation \ref{eq3}:

\begin{equation}
\resizebox{.9\hsize}{!}{$
R_{factor} = \frac{\left[\text{Im}\left(\chi_{\text{dat}}(r_{i}) - \chi_{\text{th}}(r_{i})\right)\right]^2 + \left[\text{Re}\left(\chi_{\text{dat}}(r_{i}) - \chi_{\text{th}}(r_{i})\right)\right]^2}{\left[\text{Im}\left(\chi_{\text{dat}}(r_i)\right)\right]^2 + \left[\text{Re}\left(\chi_{\text{dat}}(r_i)\right)\right]^2}
$}
\label{eq3}
\end{equation}

\noindent

Here, ${\chi_{dat}}$ and ${\chi_{th}}$ represent experimental and theoretical values, respectively, while \textit{Im} and \textit{Re} denote the imaginary and real parts, respectively. The ATOMS and ARTEMIS subroutines within the Demeter software package were employed for generating theoretical routes from crystallographic structures and fitting experimental data to theoretical simulations, respectively.

The simultaneous fitting of Fe and Ru edges could only be achieved under the assumption of Fe/Ru mixing, confirming disorder in the examined compound based on the XRD data. The best-fit theoretical spectra and experimental data are presented in Fig.~\ref{EXAFS_Fitted}, and the corresponding parameters are listed in Table~\ref{Tab:EXAFS}. The EXAFS estimates a ratio of Fe/Ru mixing equal to 59:41 and 45:55, by measuring from the Fe and Ru edges, respectively. The EXAFS data fittings align well with a disordered structure of Type-1, corroborating the XRD data analysis.

\subsection{\label{sec:Magnetism}Magnetic properties}

\begin{figure}[h]
\centerline{\includegraphics[width=.48\textwidth]{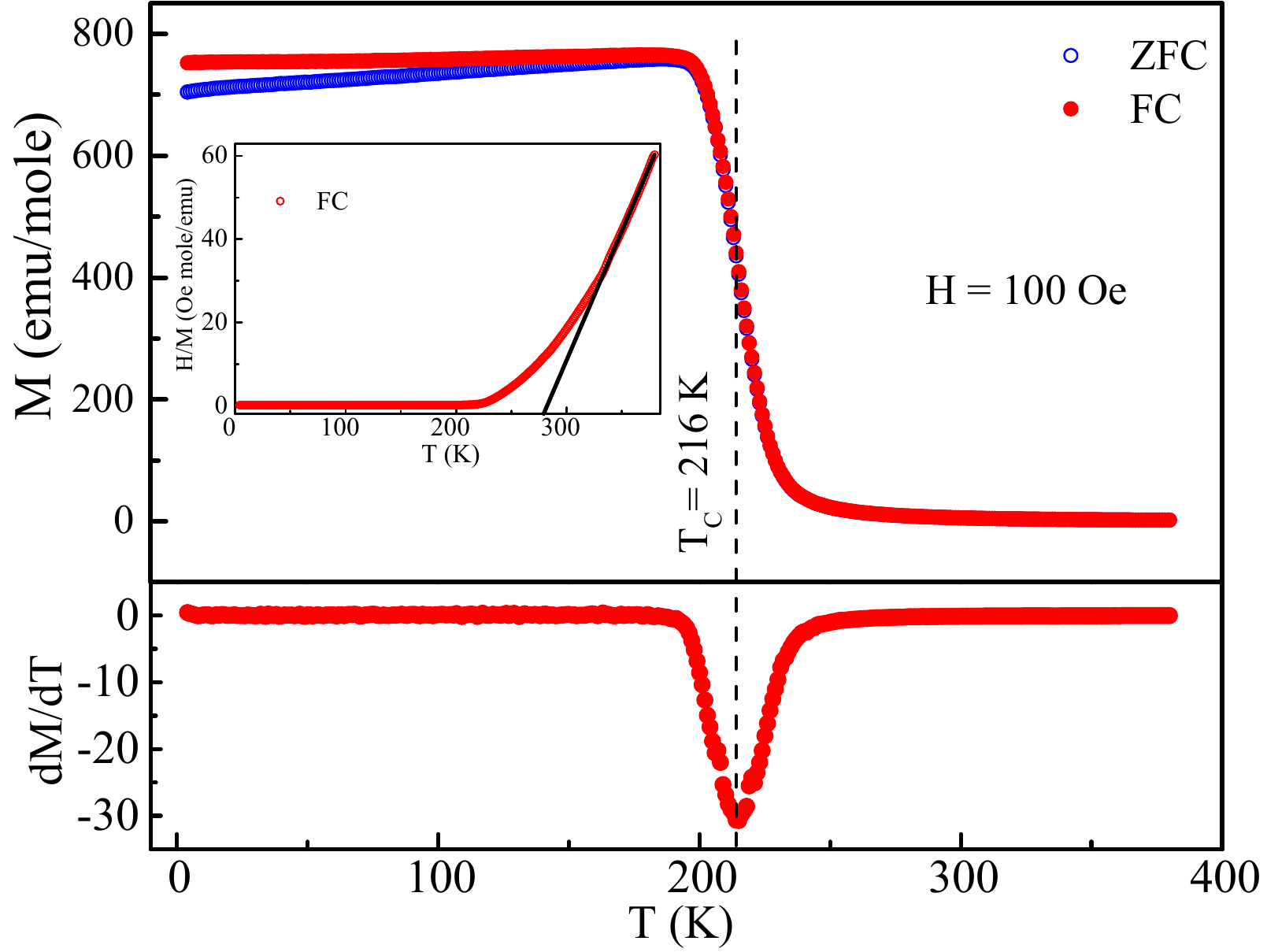}}
{\caption{Temperature dependence of magnetization in FeRuCrSi measured in 100\,Oe applied magnetic field under ZFC (blue circle) and FC (red circle) conditions. Curie temperature, $T_{\rm C}$, is determined from the minima in d$M$/d$T$ \textit{vs.} $T$ plot. Inset shows inverse susceptibility data measured under FC condition. A Curie-Weiss fit in the high temperature region with extrapolated section is also shown.}\label{MT_Fig}}
\end{figure}

Fig.~\ref{MT_Fig} represents the magnetization \textit{vs.} temperature of FeRuCrSi measured in both zero field cooled (ZFC) and field cooled (FC) condition under an application of a 100\,Oe field. Thermo-magnetic irreversibility observed between ZFC and FC magnetization vanishes when the external field strength is enhanced to 500 Oe and beyond. The susceptibility shows a sharp ferromagnetic to paramagnetic transition near its Curie temperature, $T_{\rm C}$ = 216\,K, which is determined from the minima of d($M$)/d$T$ curve. 

\begin{figure}[h]
\centerline{\includegraphics[width=.48\textwidth]{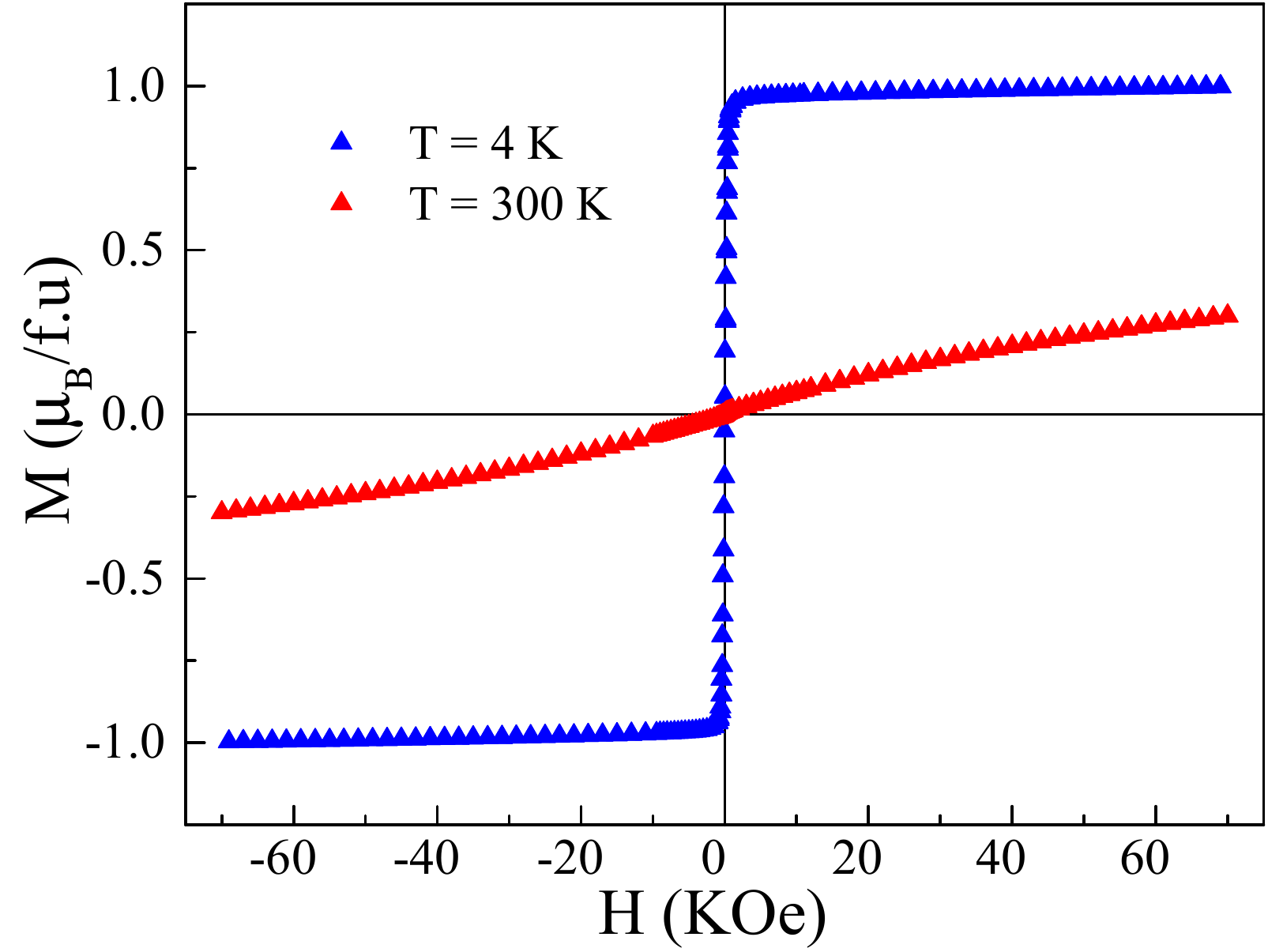}}
{\caption{Isothermal magnetization of FeRuCrSi measured at 4\,K and 300\,K.}\label{MH_Fig}}
\end{figure}

The isothermal magnetization measurements were also carried out below (4\,K) and above (300\,K) the Curie temperature of the material (Fig.~\ref{MH_Fig}). The magnetic isotherm at 4\,K exhibits a soft ferromagnetic character with the saturation magnetic moment as 0.97\,${\mu_B}$/f.u., much smaller than 2\,${\mu_B}$/f.u., expected from the S-P rule described earlier (Sec.~\ref{sec:Electronic_Structure_Ordered}). Interestingly, although the Curie temperature in our system is 216\,K, the magnetic isotherm at a much higher temperature, 300\,K, exhibits a small but definite nonlinear character, suggesting possible presence of ferromagnetic interaction at such high temperature above $T_{\rm C}$. The presence of ferromagnetic interaction, extending upto 300\,K and even beyond, has also cast its signature in the magnetic susceptibility results as the inverse susceptibility deviates from linearity between 216-330\,K (inset of Fig.~\ref{MT_Fig}). The linear character of inverse susceptibility is the hallmark of the paramagnetic system as suggested by Curie-Weiss law, and its deviation from linearity below 330\,K suggests the presence of ferromagnetic clusters persisting in the system at much higher temperature in comparison to its Curie temperature. It may be noted that in the literature, a range of Curie temperature ($T_{\rm C}$) and saturation magnetic moment ($M_{sat}$), from 346\,K to 370\,K, and 1.27 ${\mu_B}$/f.u. to 1.84 ${\mu_B}$/f.u., respectively, have been reported by various groups while investigating the same composition~\cite{matsuda2005magnetic,zhang2023atomic}. It is thus quite possible that variation of local disorder in those reported systems are responsible for such variation of $T_{\rm C}$ and $M_{sat}$. In our case, although the bulk $T_{\rm C}$ is $\sim$216\,K, it is thus likely that random distribution of atomic disorder results in some local regions having suitable conditions for higher $T_{\rm C}$ in isolated clusters.

\subsection{\label{sec:Mossbauer}M\"{o}ssbauer spectrometry}

\begin{table*}[ht]
\addtolength{\tabcolsep}{3.0pt}
\caption{Refined values of hyperfine parameters for M\"{o}ssbauer spectra measured at 77\,K and 300\,K in the FeRuCrSi Heusler alloy: Isomer shift ($\delta$) (quoted relative to $\alpha$-Fe at 300\,K), quadrupolar splitting ($\Delta$), quadrupolar shift (${2\varepsilon}$), hyperfine field (B$_{hf}$) and relative absorption areas (RA).}

\begin{tabular}{llllll}
\hline\hline
$T$ (K) & $\delta$ (mm/s) & $\Delta$ or ${2\varepsilon}$ (mm/s) & $B_{hf}$ (T) & Distribution & RA \\ 
            & $\pm$0.01       &  $\pm$0.01                          &  $\pm$0.5        &              & $\pm$1  \\ \hline\hline
            & 0.21         & 0.03     & 27.1 & 8 Fe + 0 Ru  & 12              \\ \hline
            & 0.18         & 0        & 24.1 & 7 Fe + 1 Ru  & 17              \\ \hline
            & 0.12         & 0        & 21.6 & 6 Fe + 2 Ru  & 15              \\ \hline
            & 0.20         & -0.07    & 18.8 & 5 Fe + 3 Ru  & 14              \\ \hline
77          & 0.22         & 0        & 15.3 & 4 Fe + 4 Ru  & 12              \\ \hline
            & 0.23         & 0        & 12.2 & 3 Fe + 5 Ru  & 11              \\ \hline
            & 0.24         & 0        & 5.5  & 2 Fe + 6 Ru  & 10              \\ \hline
            & 0.26         & 0.45     & -    & 1 Fe + 7 Ru  & 9               \\ \hline\hline
            & 0.025        & 0        & 22.6 & 8 Fe + 0 Ru  & 11              \\ \hline
            & 0.04         & 0        & 16.6 & 7 Fe + 1 Ru  & 18              \\ \hline
300        & 0.08         & 0         & 10.9 & 6 Fe + 2 Ru  & 15              \\ \hline
            & 0.10         & 0        & 6.6  & 5 Fe + 3 Ru  & 14              \\ \hline
            & 0.155        & 0        & -    &              & 42              \\ \hline\hline
\end{tabular}
\label{Mossbauer_Table}
\end{table*}

\begin{figure}[]
\centerline{\includegraphics[width=.48\textwidth]{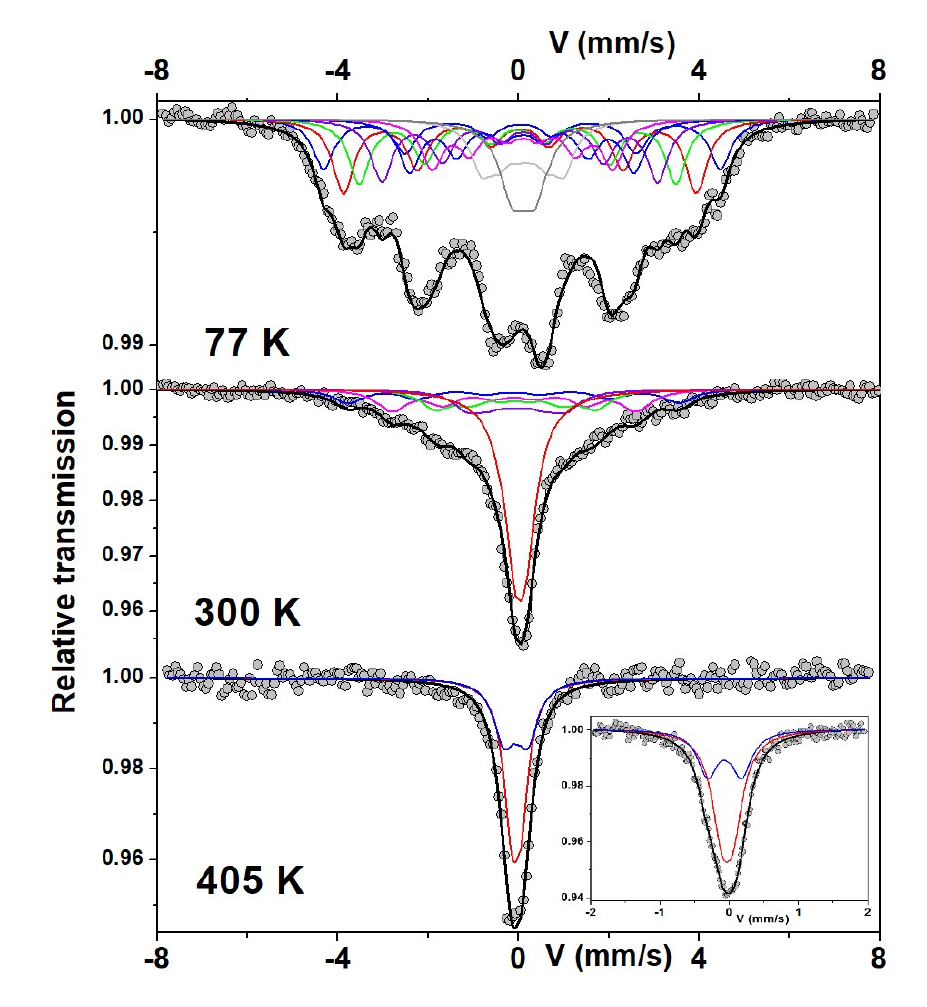}}
{\caption{M\"{o}ssbauer spectra and the multi-component description (see text) of FeRuCrSi taken at 77\,K (top), 300\,K (middle) and 405\,K (bottom). Inset shows zoomed view of the 405\,K spectra near origin.}\label{Fig_Mossbauer}}
\end{figure}

\begin{figure}[]
\centerline{\includegraphics[width=.48\textwidth]{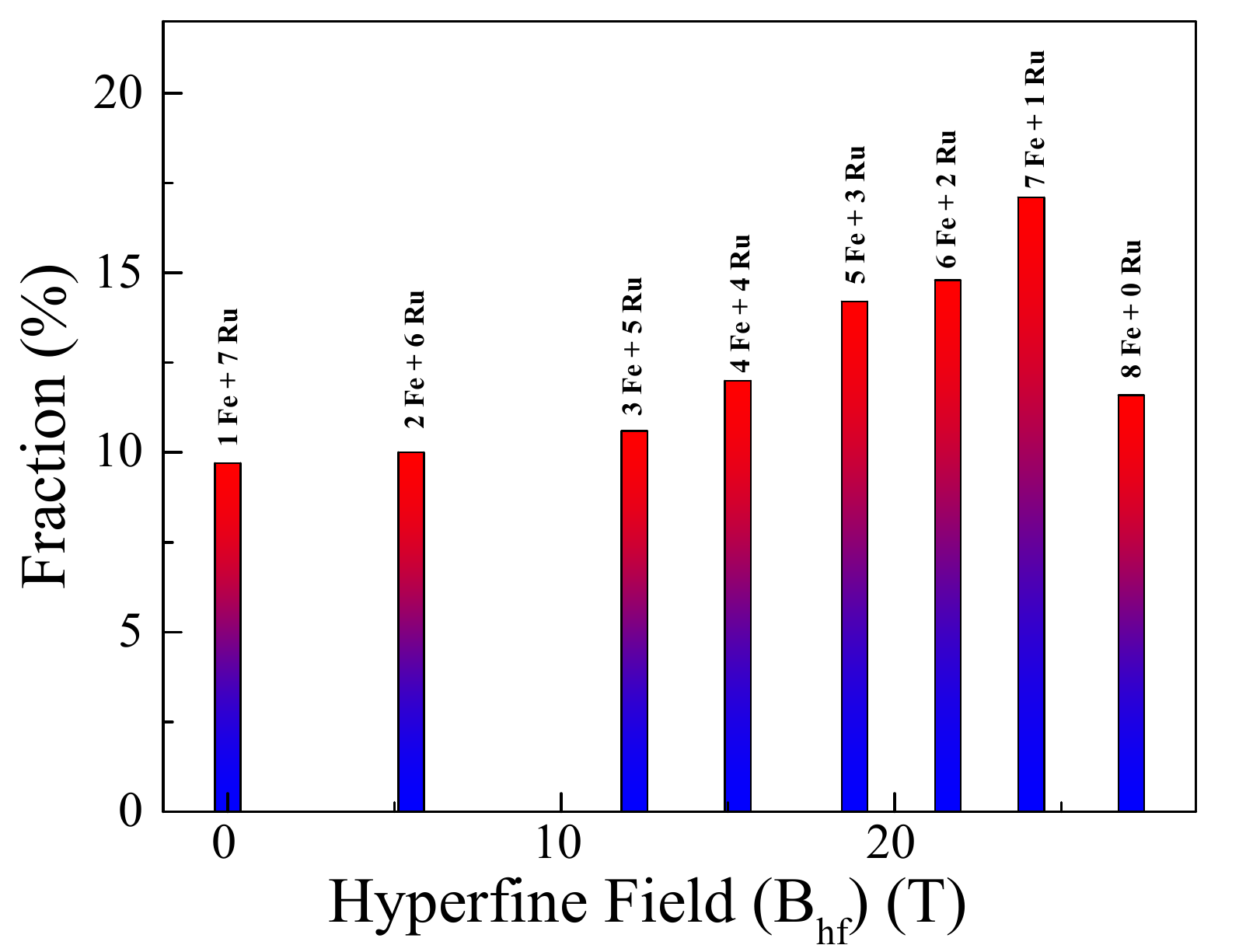}}
{\caption{Relative fraction as a function of hyperfine field for each distribution in disordered structure of FeRuCrSi at 77 K. The description excludes the configuration where all the 4\textit{c} and 4\textit{d} positions are occupied by Ru atoms only. }\label{Fig_Mossbauer_Distribution}}
\end{figure}

To verify the existence of structural disorder and its influence on the magnetic properties of FeRuCrSi, $^{57}$Fe M\"{o}ssbauer spectrometry was carried at different temperatures in the range 77\,-405\,K. The Mössbauer spectra for all temperatures (T $<$ 405 K) reveal multiple magnetic sextets superimposed on a central quadrupolar feature. Fig.~\ref{Fig_Mossbauer} clearly shows that the magnetic contribution is very prominent in the spectrum at 77\,K, whereas it is much weaker at 300\,K. Given that our complete XRD and EXAFS analyses validate a volume average of 50:50 site distribution between Fe and Ru at the 4\textit{c} and 4\textit{d} positions, one would therefore expect the M\"{o}ssbauer spectra to display only two magnetic sextets (associated with positions 4\textit{c} and 4\textit{d}) well below the Curie temperature. However, the M\"{o}ssbauer spectrum taken at 77\,K cannot be adequately described with such a simple 2-component model. To account physically for this complex hyperfine structure, a minimum of 8 components is required for a satisfactory fit, in agreement with the number of possible environments. The M\"{o}ssbauer spectrum at 77\,K (Fig.~\ref{Fig_Mossbauer}) is fitted with 7 magnetic sextets, all showing almost similar isomer shift values, and a quadrupolar component (see Table~\ref{Mossbauer_Table}), the line widths being fitted with the same value. Relative fraction as a function of hyperfine field for each distribution in disordered structure of FeRuCrSi is also presented Fig.~\ref{Fig_Mossbauer_Distribution}. The small variation of isomer shift could be taken as an indication of variations in the electron density at the Fe nuclei as a function of their atomic environments. The slightly larger isomer-shift align with the various possibilities of atomic environments sharing the same number for each species. In addition, the isomer shift follows a more or less regular increase as the hyperfine field decreases, which includes the quadrupolar component. Another possible description which can be obtained by considering a linear decrease of isomer shift correlated to decreasing hyperfine field values, leads to fairly similar conclusions.   

The refined values of the hyperfine parameters are given in Table~\ref{Mossbauer_Table}. The average hyperfine field is thus estimated at 16.8\,Tesla for such an 8-component fit comprising 7 different magnetic sextets. For the calibration of magnetic moment from hyperfine field values, usually $\alpha$-Fe with a magnetic moment of 2 ${\mu_B}$ producing a hyperfine field of 33 Tesla is considered as standard. So, for a 16.8 T magnetic field, a moment of 1.0\,${\mu_B}$ is obtained, which matches very closely with the moment obtained from isothermal magnetization (Sec.~\ref{sec:Magnetism}). This is a very surprising result that states that Fe atoms are the primary source to the magnetic moment : indeed, this convincing result is completely inconsistent with all the reported DFT calculations~\cite{guo2018magnetic,zhang2023atomic}, as well as with the one presented in this work (Sec.~\ref{sec:Electronic_Structure_Ordered}), which predicts that Cr should be the major source to the magnetic moment, while the magnetic moment at the Fe site would be induced in nature. This is also one of the main reasons why the isothermal saturation magnetization values for this compound deviate from those proposed by the DFT calculations. This also explains why with increasing Fe concentration, the saturation magnetic moment and $T_{\rm C}$ in Fe$_x$Ru$_{2-x}$CrSi have been reported to increase linearly~\cite{matsuda2005magnetic} and not remain constant to 2 ${\mu_B}$/f.u. Fe being the major contributor of magnetic moment in this series also explain why this compound experimentally deviates from the theoretically predicted HMF character.

\begin{figure}[]
\centerline{\includegraphics[width=.48\textwidth]{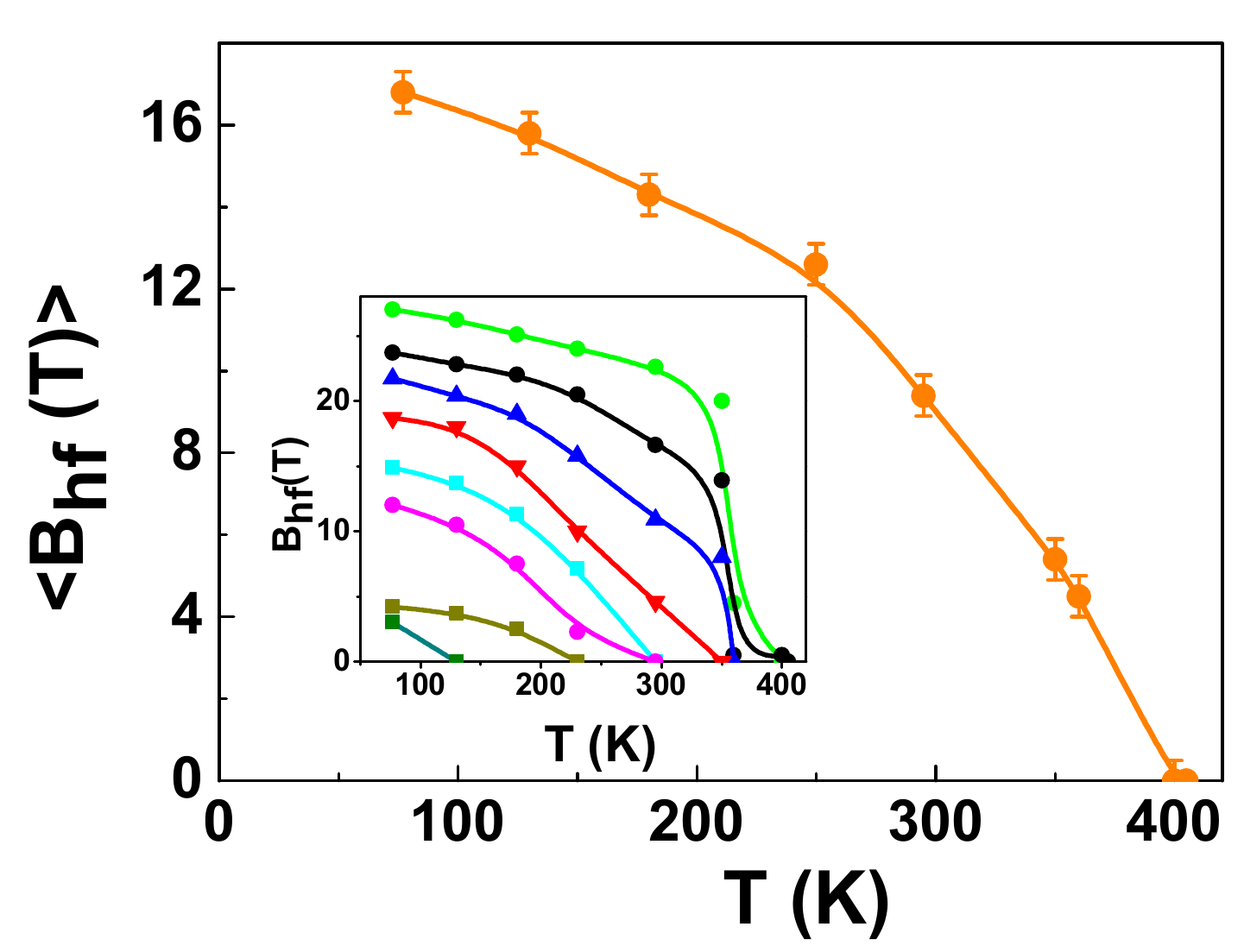}}
{\caption{Evolution, as a function of temperature, of the mean values of the hyperfine field estimated from the M\"{o}ssbauer spectra on the FeRuCrSi powder sample. The inset shows the evolution of the hyperfine field corresponding to the different individual magnetic components. Solid lines are guide to the eye. }\label{Fig_9}}
\end{figure}

It is interesting to note, as discussed above, that the $^{57}$Fe M\"{o}ssbauer spectra could only be analysed by considering 8 different components corresponding to 8 different Fe environments, whereas the XRD and EXAFS measurements suggest that Fe atoms are randomly distributed along Ru atoms only in two crystallographic sites, i.e. 4\textit{c} and 4\textit{d}. This apparent contradiction can be easily understood if we consider that the random distribution of Fe atoms resulting from a large number of different localised atomic environments. 
When Fe is randomly distributed between 4\textit{c} and 4\textit{d} sites, there can be 9 different possibilities of atomic distributions, ranging from all 8 atoms being Fe in some unit cells, all 8 Ru in other unit cells, and the rest of the unit cells has varying degree of mixing with the restriction that when averaged over all unit cells, the composition remains FeRuCrSi. These 9 possibilities are: 8 Fe + 0 Ru, 7 Fe + 1 Ru, 6 Fe + 2 Ru, 5 Fe + 3 Ru, 4 Fe + 4 Ru, 3 Fe + 5 Ru, 2 Fe + 6 Ru, 7 Fe + 1 Ru, 0 Fe + 8 Ru. Since $^{57}$Fe M\"{o}ssbauer measurements can only detect Fe atoms, the configuration involving 0 Fe + 8 Ru would not be related to the M\"{o}ssbauer results, but would provide a very useful nature about the 8 remaining configurations. 

\begin{figure}[]
\centerline{\includegraphics[width=.48\textwidth]{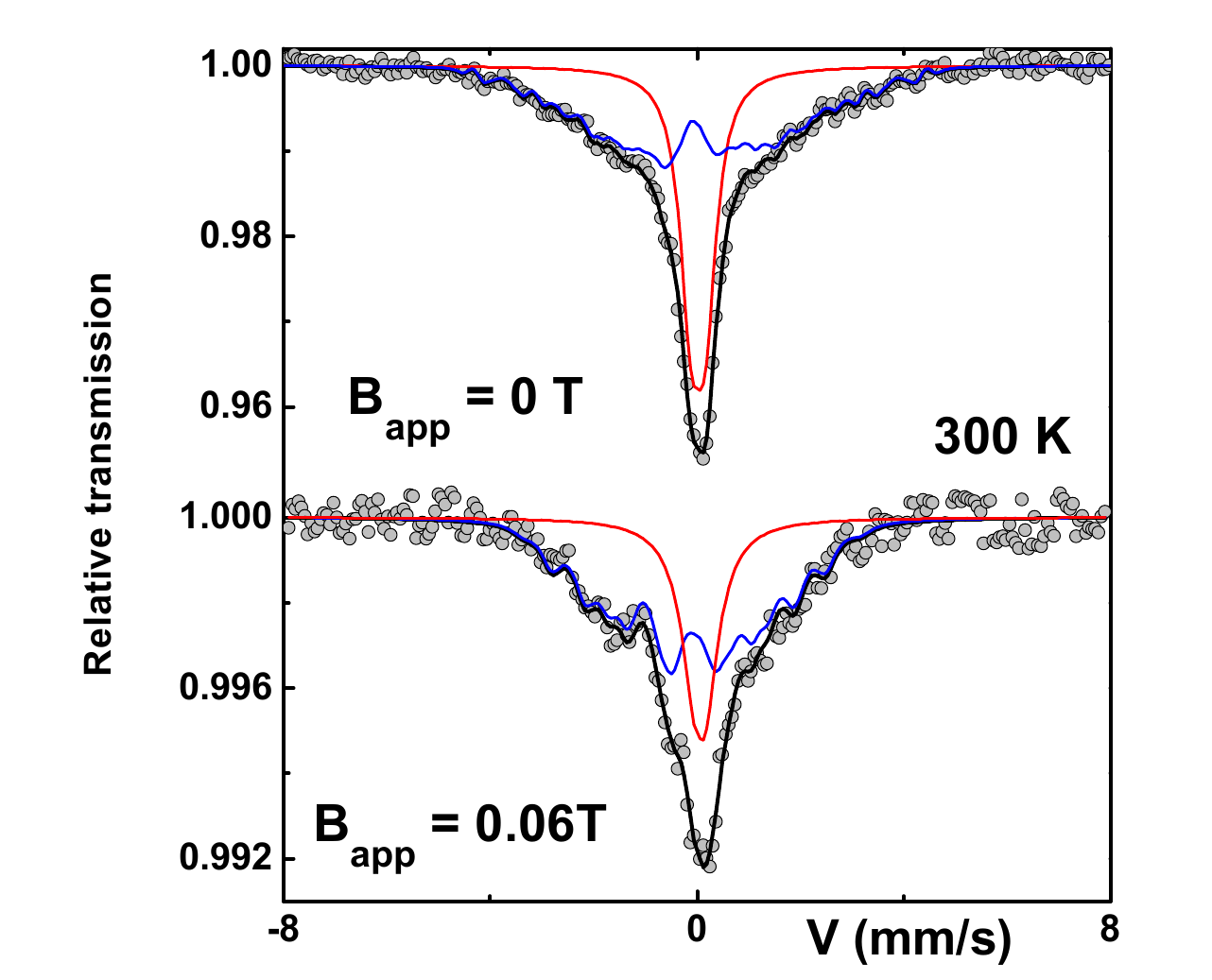}}
{\caption{M\"{o}ssbauer spectra at 300K obtained on the FeRuCrSi powder sample without magnetic field (top) and with an external magnetic field of approximately 0.06T (bottom), oriented perpendicular to the $\gamma$-beam). The red and blue lines correspond to the paramagnetic and magnetic components, respectively.}\label{Fig_10}}
\end{figure}

As it has been previously mentioned that Fe$_2$CrSi exhibits a ferromagnetic ordering at 520\,K~\cite{luo2007electronic}, while all other members of the Ru-diluted Fe$_x$Ru$_{2-x}$CrSi series possess a lower Curie temperature~\cite{hiroi2009magnetic,hiroi2009magnetic,PhysRevB.103.094428}, an attempt can be made to correlate the reduction of $T_{\rm C}$ in this series with a reduced magnetic interaction strength caused by non-magnetic (Ru) dilution of the local atomic environment of Fe. Using the same logic in our system, we correlated the different hyperfine magnetic field values of the 8 components of the M\"{o}ssbauer spectra at 77\,K with 8 different atomic environments of the Fe atoms. 
The largest hyperfine field is shown by the Fe atoms in the unit cell that has only Fe atoms in all the 4\textit{c} and 4\textit{d} positions, while the single quadrupolar component which shows no hyperfine field has only 1 Fe atom together with 7 Ru atoms in the unit cell. Excluding the 0 Fe + 8 Ru configuration, our M\"{o}ssbauer analysis suggests that all the other 8 configurations are present in our system with almost equal probabilities (Table~\ref{Mossbauer_Table} and Fig.~\ref{Fig_Mossbauer_Distribution}). 
The average value of the magnetic hyperfine field represents the average magnetic interaction that in turn determines the bulk saturation moment as well as $T_{\rm C}$. This is also corroborated from our M\"{o}ssbauer spectrum taken at 300 K, where the first four components presented in Table~\ref{Mossbauer_Table} and inset of Fig.~\ref{Fig_9} still show weak hyperfine fields, although the bulk \(T_C\) of our sample, as determined from magnetic susceptibility measurements, is much lower (\(\sim 216~\mathrm{K}\)). A series of M\"{o}ssbauer spectra were recorded at different temperatures from 77 K until a pure paramagnetic spectrum was obtained at 405 K, as observed in Fig.~\ref{Fig_Mossbauer}. It is concluded that the M\"{o}ssbauer spectra recorded in the 150 K–400 K range result from the presence of magnetic sextets and a central quadrupolar component. They were fitted using some magnetic components and a quadrupolar doublet, but the values of some hyperfine parameters have been fixed during the fitting procedure to compensate for the lack of resolution of the central part, illustrated by the broadening of the central component. The mean value of the characteristic hyperfine field for the whole sample is given in Fig.~\ref{Fig_9}. In addition, the values of the hyperfine field for the different components have also been plotted as a function of temperature (Fig.~\ref{Fig_9}: inset). This description confirms the progressive temperature dependence of the magnetically ordered–paramagnetic transition of the Fe moments, which is fully achieved at 405 K. 

To better understand the origin of this phenomenon, we compared the hyperfine structures as obtained at 300 K on the powder sample without magnetic field and with an external magnetic field of approximately \(0.06~\mathrm{T}\) (as shown in Fig.~\ref{Fig_10}), oriented perpendicular to the direction of the \(\gamma\)-beam produced by a simple magnet placed far from the source (approximately 20 cm) to avoid the presence of polarization effects. The magnetic component proportion increases under the effect of the external magnetic field, which clearly demonstrates the presence of a progressive blocking of the Fe magnetic moments, i.e., the presence of certain dynamic effects of magnetic clusters. The non-linearity of the inverse susceptibility (Fig.~\ref{MT_Fig}: inset) below $\sim$330\,K is also caused by the presence of such magnetic clusters in our system. The different distribution of the local atomic environment may also explain the variation in $M_{sat}$ and $T_{\rm C}$  reported by various groups for the same equiatomic quaternary Heusler alloy FeRuCrSi. The presence of larger fractions of configurations with more Fe magnetic moments would lead to larger $M_{sat}$ and $T_{\rm C}$ and vice versa. 

Our results therefore establish a rather rare and complex scenario in which the main physical properties are strongly correlated to the material chemistry in the form of random atomic disorder on a localised scale. At this stage, it is difficult, on the basis of all the current data, to specify accurately the nature and size of these local atomic environments that lead to a discrete distribution of hyperfine fields, given the number of possible environments associated with the presence of 4 different atomic species. This level of understanding of atomic arrangement using M\"{o}ssbauer spectrometry is remarkable and echoes previous studies based on another related technique, namely nuclear magnetic resonance measurements in other Heusler alloys~\cite{gupta2022coexisting,wurmehl201355,wurmehl2007probing}.

\subsection{\label{sec:Electronic_Structure_Disordered} Electronic structure calculations for disordered crystal structure}

\begin{figure}[]
\centerline{\includegraphics[width=.48\textwidth]{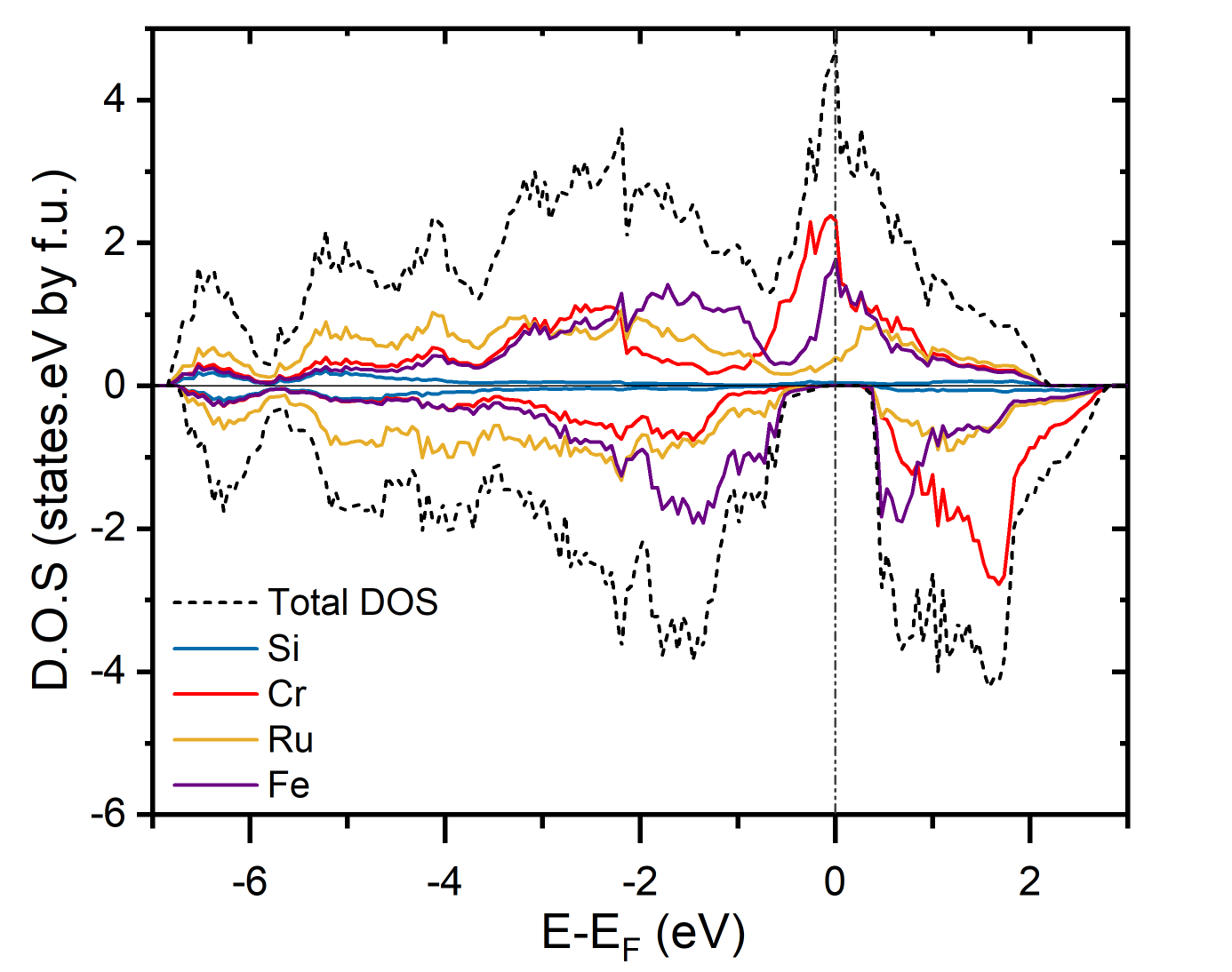}}
{\caption{Partial density of states for FeRuCrSi for Type-1 disordered structure between Fe and Ru atoms.}\label{Fig_DOS_Disorder}}
\end{figure}

Following the XRD measurements and a possible 50:50 site disorder between Fe and Ru atoms (Type-1), additional calculations have been made on FeRuCrSi to explore its electronic structure, considering the observed disorder by the theoretical analysis (Fig.~\ref{Fig_DOS_Disorder}).
The DFT calculation show that although the disorder slightly stabilizes the structure, the spin polarization is not affected (system maintains \textit{P}= 100$\%$) and the total magnetic moment remains equal to 1.94\,$\mu_B$, consistent with both the ordered structure and the Slater-Pauling value. 

However, the distribution of magnetic moments on each atom in Type-1 (both ordered and disordered) does not agree with the experimental measurements regarding the magnetic ordering. In fact, the DFT calculations for Type-1 show that Cr is the main contributor to the magnetic moment. This is contradicted by the M\"{o}ssbauer spectrometry and magnetic measurements, which indicate that Fe should be the main contributor to the magnetic moment.
This kind of magnetic ordering was reproduced by a Type-3 configuration, with Cr and Ru in a separate plane from Fe (in both ordered and disordered configurations). However, Type-3 is calculated to be slightly less stable than Type-1 at 0 K, but with Fe having a magnetic moment of approximately 2 $\mu_B$, antiparallel to the -1 $\mu_B$ carried by Cr. This configuration is relatively more consistent with the observations reported in Fig.~\ref{XRD}.

The differences between experimental results and theoretical predictions can be attributed to several factors, as the calculations were performed at 0 K on so-called ideal structures. The disordered Type-3 configuration, influenced by temperature or synthesis conditions, could therefore be favored over Type-1, even though structural measurements do not provide a clear distinction. Given all these experimental and theoretical results, no firm conclusion can be drawn on the HMF state of this compound as well as in the entire Fe$_x$Ru$_{2-x}$CrSi series.

\subsection{\label{sec:Conclusion}Conclusion}

As a conclusion, this work utilized X-ray diffraction, extended X-ray absorption fine structure, $^{57}$Fe M\"{o}sbauer spectrometry and magnetic measurements in a quaternary Heusler alloy, FeRuCrSi, belonging to the Fe$_x$Ru$_{2-x}$CrSi series. These techniques led us to conclude that FeRuCrSi crystallizes in a Type-1 structure with 50:50 site disorder between the Fe and Ru atoms. DFT calculations show this last structure to be the most stable. However, these calculations also indicate that Cr would be the main contributor to the magnetic moment and predict a saturation magnetic moment of 2\,$\mu_B$/f.u., with robust half-metallicity (100\% spin polarization). These theoretical predictions are contradicted by experimental results, showing that Fe is the main contributor to magnetism (saturation moment of 0.97\,$\mu_B$/f.u.) and challenging the half-metal character. According to DFT calculations, this discrepancy could have been resolved if FeRuCrSi had crystallized in a different structure, where Cr and Ru are placed in a separate plane from Fe (Type--3). However, this arrangement is slightly less stable than Type-1 at 0 K. Despite identical compositions with a mixing of Ru and Fe, there are reports of significant variations in Curie temperatures and saturation moments, which often deviate from theoretical predictions. The dominant role of Fe in the magnetic properties of the sample explains the experimental deviations from theoretical predictions and highlights the impact of the local atomic disorder on the physical properties. The work presented here highlights the complex inter relationship between the chemistry of the material, atomic disorder, and the physical properties of the Fe$_x$Ru$_{2-x}$CrSi series. 

\vspace{1cm}
\centerline{\textbf{Acknowledgement}}
S.G and S.C would like to sincerely acknowledge SINP, India and UGC, India, respectively, for their fellowship. DFT calculations were performed using HPC resources from GENCI-CINES (Grant 2021-A0100906175). Part of this work was performed under CSRP project 6908-3 of the Indo-French Centre for Promotion of Advanced Research, New Delhi, India.

\bibliographystyle{apsrev4-2}
%

\end{document}